# Chaos and dynamics of spinning particles in Kerr spacetime

**Wenbiao Han**

**Abstract** We study chaos dynamics of spinning particles in Kerr spacetime of rotating black holes use the Papapetrou equations by numerical integration. Because of spin, this system exists many chaos solutions, and exhibits some exceptional dynamic character. We investigate the relations between the orbits chaos and the spin magnitude *S*, pericenter, polar angle and Kerr rotation parameter *a* by means of a kind of brand new Fast Lyapulov Indicator (FLI) which is defined in general relativity. The classical definition of Lyapulov exponent (LE) perhaps fails in curve spacetime. And we emphasize that the Poincaré sections cannot be used to detect chaos for this case. Via calculations, some new interesting conclusions are found: though chaos is easier to emerge with bigger S, but not always depends on *S* monotonically; the Kerr parameter *a* has a contrary action on the chaos occurrence. Furthermore, the spin of particles can destroy the symmetry of the orbits about the equatorial plane. And for some special initial conditions, the orbits have equilibrium points.

## 1 Introduction

In the last decade or so, chaos dynamics in general relativity began to be widely appreciated [1] against to in classical dynamics has been researched about 100 years from Poincaré [2]. Because general relativity is a nonlinear theory, so chaos is often visible. There are two main aspects in chaos dynamics in general relativity. One is the dynamics evolution of gravitational field itself, such as cosmology models especially Mixmaster Universe model [3–6]. Another is the test particles' dynamical character

W. Han (✉)
Shanghai Astronomical Observatory, Graduate School of the Chinese Academy of Sciences,
80 Nandan Road, Shanghai 200030, China
e-mail: wbhan@shao.ac.cn

in known metrics. Some interesting models are integrable in Newtonian theory but exhibiting chaos in general relativity, such as the case of two fixed black holes [4,7] and the Schwarzschild black hole plus a dipolar shell [8,9]. Letelier et al. also discussed chaos dynamics of a test particle in slow rotating black holes with dipole halo [10] and the perturbed Schwarzschild spacetime [11]. Many references focus on the geodesic motion of particles in superposed Weyl fields [12–14], a Schwarzschild or Kerr black hole with multipole halos. On the other hand, Suzuki and Maeda studied a spinning particle in Schwarzschild and Kerr spacetime [15,16], Hartl investigated the last one in detail [17,18].

In the recent years, because of the detecting project of gravitational-wave, such as Laser Interferometer Space Antenna (LISA), Laser Interferometer Gravitational Wave Observatory (LIGO) and Astrodynamical Space Test of Relativity using Optical Devices (ASTROD), the spinning compact binaries dynamics attracted people great interesting [19–25], since the gravitational-wave detection cannot succeed when chaos is present. So, the study of chaos dynamics of spinning particles in Schwarzschild and Kerr spacetime called great interest.

Spinning particles in Kerr spacetime, just a strong gravity central body, like a Kerr black hole or neutron star, is a basic model in cosmos, has important physics significance. Meanwhile, this model is an important source of gravitational wave. Because the orbits of the spinning particles are likely chaotic, the research of this model is very interesting.

All we know, nonspinning particles' orbits are regular in Schwarzschild or Kerr spacetime. But for spinning particles, the orbits perhaps are chaotic. Corinaldesi and Papapetrou first discussed the motion of spinning test particle in Schwarzschild spacetime [26]. The Kerr or Kerr–Newman spacetime case was also analyzed by some researchers [27–32]. In [33,34] the gravitational waves produced by a spinning particle falling into a Kerr black hole or moving circularly around it were discussed and the energy emission rate from those systems was calculated. Suzuki and Meada investigated more generic motion of a spinning particle around a Schwarzschild black hole and pointed out that the spin effect can make some orbit chaotic [15]. Furthermore, they studied the innermost stable circular orbit of a spinning particles in Kerr spacetime [16], and they investigated orbit chaos by Poincaré sections. Then, Kiuchi and Maeda researched the gravitational waves from the above dynamical system [35]. They also analyzed gravitational waves from spinning particles around Schwarzschild black hole [36].

On the other hand, Michael Hartl discussed the chaos dynamics of spinning test particles in Kerr spacetime by means of the classical definition of Lyapunov exponent (LE), and find that chaos appears only for physically unrealistic values of the spin parameter firstly [17]. More than of that, he did a detailed survey of spinning test particle orbits in Kerr spacetime by use of same tool [18].

In general, the methods for quantifying the ordered or chaotic nature of orbits in general relativity usually are Poincaré sections, Lyapunov exponent and fast Lyapunov indicator (FLI). If the system's degrees of freedom are not more than four, Poincaré sections is one of the most common qualitative tools. Karas and Vokrouhlicky firstly introduced LE to study the motion near a black hole in general relativity [37]. But LE and FLI of classical dynamics sometimes invalidate while being used in general relativity. So Wu et al. gave definitions of LE and FLI with two nearby trajectories in curved spacetime [38,39]. There is a close relation between the FLI and LE: the FLI divided by time *t* tends to the LE when the time is sufficiently large. Besides, overflow of the lengths of tangential vectors in the case of a chaotic orbit can be avoided because the integration time is not long enough. This is the reason why the indicator is classified as a "fast" method.

Based on the reason above mentioned, in this paper, we continue to study the chaos and dynamics of spinning particles around a central Kerr black hole by the brand new FLI in general relativity in stead of classical LE used by Hartl. We emphasize that, in this letter, because of too many degrees of freedom, the Poincaré sections are invalid in detecting chaos, but can be used to study the dynamical structure. In our case, the FLI in general relativity is a valid and convenient method for detecting chaos. So, by use of it, we research the dependence of orbit chaos on some dynamical parameters: initial radii, polar angle, Kerr and spin parameter. We find that the chaotic degree does not monotone increase with the spinning magnitude, and the Kerr parameter appears to counteract chaos. Furthermore, we also find the spin of particles can destroy the symmetry of the orbits about the equatorial plane. And some strange dynamical structures emerge in special initial conditions.

The paper is organized as follows. The basic equations for spinning particles in curve space are reviewed briefly at fist, and the numerical integration method is discussed in Sect. 3. Then, the FLI of Wu is introduced simply in Sect. 4, and we explain why we choose FLI but not Poincaré section at the next section. In Sect. 6, we study the relations of chaos with different parameters. Then we give some strange dynamical character in Sect. 7, meanwhile we investigate the dependence of the maximal inclinations with spin parameter. At the end, we present a simple summary and discussion.

## 2 Papapetrou equations for spinning test particles

The equations of motion of a spinning test particle in a curve spacetime were given first by Papapetrou [40], and then reformulated a set of more clear form by Dixon [41]. In this paper, we measure all times and lengths in terms of $M$, measure the momentum of the particle in terms of $\mu$, and measure the spin in units of $\mu M$. The $\mu$ represents the rest mass of particle, and $M$ is the mass of center massive body. So these equations are [42],

$$\frac{dx^\mu}{d\tau} = \upsilon^\mu$$
$$\frac{dp^\mu}{d\tau} = -\left(\frac{1}{2} R^\mu_{\nu\rho\sigma} S^{\rho\sigma} + \Gamma^\mu_{\nu\sigma} p^\sigma\right) \upsilon^\nu \qquad (1)$$
$$\frac{dS^{\mu\nu}}{d\tau} = 2(p^{[\mu} \upsilon^{\nu]} + \Gamma^{[\mu}_{\rho\sigma} S^{\nu]\rho} \upsilon^\sigma),$$

where, $\upsilon^\mu$ is the four-velocity, or the tangent to the particle's worldline. $p^\mu$ and $S^{\mu\nu}$ are the momentum and the spin tensor respectively. Here, $\tau$ is defined as proper time.

The $S^{\mu\nu}$ is a anti-symmetry tensor, defined by the particle's stress–energy tensor $T^{\mu\nu}$,

$$S^{\mu\nu} = \int d^3x (x^\mu T^{\nu 0} - x^\nu T^{\mu 0}). \tag{2}$$

The Eq. (1) is called the pole–dipole approximation, where the multipole moments of the particle higher than mass monopole and spin dipole are ignored.

Because of spinning, the motion of particle does not follow the geodesic, so the $p^\mu$ is no longer parallel to $\upsilon^\mu$. If following Dixon, choosing the rest frame of the particle's center of mass, we can get one of the spin supplementary conditions [41]

$$p_\mu S^{\mu\nu} = 0. \tag{3}$$

And from this equation, we can find the relation of $p^\mu$ and $\upsilon^\mu$ [42],

$$\upsilon^\mu = p^\mu + \frac{2 S^{\mu\nu} R_{\nu l \kappa \lambda} p^l S^{\kappa\lambda}}{4 + R_{\alpha\beta\gamma\delta} S^{\alpha\beta} S^{\gamma\delta}}. \tag{4}$$

For the more, there are four others constraints [17,42],

$$p^\mu p_\mu = -1 \tag{5}$$

$$S^{\mu\nu} S_{\mu\nu} = S^2 \tag{6}$$

$$E = -p_t + \frac{1}{2} g_{t\mu,\nu} S^{\mu\nu}, \tag{7}$$

and

$$J_z = p_\phi - \frac{1}{2} g_{\phi\mu,\nu} S^{\mu\nu}. \tag{8}$$

Here, the $S$ denotes the spin magnitude, and quantifies the size of the spin, so plays a crucial role in determining the behavior of spinning particle systems. The $E$ and $J_z$ represent energy and $z$ angular momentum of particle respectively. In fact, Kerr spacetime also exist another integral, Carter constant $Q$ [43,44].

## 3 Numerical integration of Papapetrou equations

In this section, we introduce our numerical methods for Eq. (1). There are 24 variables in Papapetrou equations (1), but because the anti-symmetry of spin tensor $S^{\mu\nu}$, so the number of variables decreases to 14. For the reason of existence of many constraints, the initial values cannot be assigned arbitrarily, we first appoint several initial parameters, and the rest are computed from the integrals (3), (5)–(8). For the reason of $S^{\mu\nu}$ can be deduced from 1-form spin vector, so the problem will be simpler furthermore. The tensor and vector formulations of the spin are related by

$$S_\mu = \frac{1}{2} \varepsilon_{\mu\nu\alpha\beta} u^\nu S^{\alpha\beta} \tag{9}$$

and

$$S^{\mu\nu} = -\varepsilon^{\mu\nu\alpha\beta} S_\alpha u_\beta, \qquad (10)$$

where $u_\nu = p_\nu/\mu(= p_\nu$ in our units), $\varepsilon^{\mu\nu\alpha\beta}$ is Levi–Civita tensor. In addition, the spin satisfies the condition

$$S_\mu S^\mu = \frac{1}{2} S_{\mu\nu} S^{\mu\nu} = S^2, \qquad (11)$$

and

$$p^\mu S_\mu = 0 \qquad (12)$$

So we set the initial conditions $t, r, \theta, \phi, p^r, S^r, S^\theta$, and the left $p^t, p^\theta, p^\phi, S^t$ and $S^\phi$ will be calculate by Eqs. (5), (7), (8), and (9)–(12). Then from Eq. (10), we can get the initial value of $S^{\mu\nu}$, thus the Eq. (1) can be integrate numerically. And in this paper, we integrate (10) directly, instead of the equations in Hartl' paper [17,18], because this can reduce CPU time greatly.

For convenience, $t = 0$, $p^r = 0$ and $\phi = 0$ is fixed in initial time. Because of [43]

$$\Sigma^2 \left(\frac{dr}{d\tau}\right)^2 = R(r), \qquad (13)$$

where

$$R(r) = [E(r^2 + a^2) - aL_z]^2 - \Delta[r^2 + (L_z - aE)^2 + Q], \qquad (14)$$

and we use the standard auxiliary variables

$$\Sigma = r^2 + a^2 \cos^2 \theta, \qquad (15)$$

and

$$\Delta = r^2 - 2Mr + a^2. \qquad (16)$$

$Q$ is Carter constant

$$Q = p_\theta^2 + \cos^2 \theta [a^2(m^2 - E^2) + \sin^{-2} \theta L_z^2]. \qquad (17)$$

So, if giving the initial $r_0, \theta_0$ and $p_0^\theta$, basing Eq. (13), the apocenter and pericenter can be fixed. This hints us to investigate the relation between the dynamical characters of the above-mentioned system and the parameters $r_0, \theta_0, a, S$.

Our integrator is RKF7 (8), step size is 0.1, and adopt conservations Eqs. (3) and (5)–(8) to check the numerical precision, and we find it can achieve $10^{-13}$ to $10^{-14}$ while integration time is $10^5$.

## 4 Fast Lyapulov indicator in general relativity

Lyapulov exponent (LE) is a very good tool to distinguish chaos and non-chaos in dynamical system. Sometimes it may cost too much CPU time to computer until reach convergence. So Froeschlé et al. [45,46] defined FLI which need not to be computed until to convergence. FLI not only is very fast tool to find chaos, but also can sketch out the global structure of phase space.

But the classical definitions of LE and FLI may be noneffective when they are used in general relativity. In curve spacetime, the value of classical LE or FLI depends on the choice of time and space coordinates. In other words, there would be different values of the classical LE and FLI in variable coordinate system, even fail to find chaos. So Wu et al. [38,39] give new definitions in general relativity. We would like to use invariant FLI in curved spacetime to detect chaos. Now we introduce the new FLI simply.

The definition of FLI bases on two-nearby-orbits method which simplifies the problem greatly, because we do not need to deduce complicated geodesic deviation equations. For a free particle move along the geodesic,

$$\ddot{x}^\mu = -\Gamma^\mu_{\alpha\beta} \dot{x}^\alpha \dot{x}^\beta. \tag{18}$$

Let two particles, an observer and his "neighbor," move on two nearby-trajectories in a curved spacetime. At the observer's proper time $\tau$, the observer is at the point $O$ with coordinate $x^\alpha$ and 4-velocity $v^\alpha$, and his neighbor reaches the point $\tilde{O}$ with coordinate $\tilde{x}^\alpha$, we attain the deviation vector

$$\Delta x(\tau) = \Delta x^\alpha(\tau) = \tilde{x}^\alpha(\tau) - x^\alpha(\tau). \tag{19}$$

So the distance along the Kerr geodesics of the neighbor measured by the observer at proper time $\tau$ is

$$||\Delta \mathbf{x}(\tau)||' = \sqrt{\left|\int_{\mathbf{x}_O}^{\mathbf{x}_N} g_{\mu\nu}(\mathbf{x}) dx^\mu dx^\nu\right|}, \tag{20}$$

the $\mathbf{x_O}$, $\mathbf{x}_N$ represent the 4-vectors of "observer" and "neighbor" individual. Obviously, this is a quite complex calculation. In our letter, for nonchaotic cases, $\Delta x^\mu \approx dx^\mu$; Even for chaos orbits, because we renormalize the neighbor's orbits while $\|\Delta \mathbf{x}\|$ achieving at 0.1 (will be mentioned later), so the below definition of space distance $\|\Delta \mathbf{x}(\tau)\|' \sim \|\Delta \mathbf{x}(\tau)\|$,

$$||\Delta \mathbf{x}(\tau)|| = \sqrt{|\Delta x \cdot \Delta x|} = \sqrt{|g_{\mu\nu} \Delta x^\mu \Delta x^\nu|}. \tag{21}$$

But Eq. (21) is simpler than Eq. (20), and only has tiny difference. So Wu et al. adopted the definition (21) in reference [39]. Thus we define the FLI as follows:

$$\text{FLI}(\tau) = \log_{10} \frac{||\Delta x(\tau)||}{||\Delta x(0)||}, \quad (22)$$

where $\Delta x(0)$ represents the deviation vector at initial time. The Eq. (22) just is the definition of new FLI by Wu et al. [39].

For geodesic motion, Wu et al. have proved the timelike distance (21) is regular as long as time evolvement, so need not projection operation. But in our example, the particles are no longer along the geodesic, so we must project (21) to observer local space. For the same reason referred above, then the projected space distance is

$$\Delta x(\tau) = \sqrt{h_{\alpha\beta} \Delta x^{\alpha}(\tau) \Delta x^{\beta}(\tau)}, \quad (23)$$

where, space projection operator

$$h^{\alpha\beta} = g^{\alpha\beta} + v^{\alpha} v^{\beta}. \quad (24)$$

The FLI($\tau$) in Eq. (22) cannot be computed without renormalization because the distance between the two particles would expand so fast in the case of chaos as to it could reach the chaotic boundary to cause saturation. Wu et al. [38,39] appoint that it works well when we choose the initial distance $|\Delta x(0)| \approx 10^{-7}$ to $10^{-9}$, and let $|\Delta x| = 0.1$ as the critical value to implement the renormalization. Let $k(k = 0, 1, 2, \ldots)$ be the sequential number of renormalization, and then calculate the FLI with the following expression [39]:

$$\text{FLI}_k(\tau) = -k(1 + \log_{10} ||\Delta x(0)||) + \log_{10} \frac{||\Delta x(\tau)||}{||\Delta x(0)||}. \quad (25)$$

We would use Eq. (25) to calculate FLI in next section.

## 5 Detect chaos by FLI

In this section, we study the availability of several methods in detecting chaos, and point out that FLI is work good at this. Because FLI need not to compute to convergence, just observe the increase of FLI with time. So it would decrease the computing time greatly compare to LE. We are typical to compute to $2 \times 10^4 M$, if FLI rise exponential with time, we think the system is chaos. And if FLI increase only linearly, the system is nonchaotic. In this paper, via many calculations, we find FLI = 6 is very good critical value for distinguish chaos while computing time achieving $2 \times 10^4$. When FLI is bigger than 6, we think the orbit is chaotic, else is nonchaotic.

So FLI is a very appropriate tool to detect chaos. We emphasize to point out that Poincaré sections cannot distinguish chaos in our example, because the system have seven dimensions. But Poincaré sections can help us understand the dynamical structure. Obviously, we also cannot distinguish chaos from the particles trajectory. From Figs. 1, 2, 3 and 4 show the validity of all kinds method.

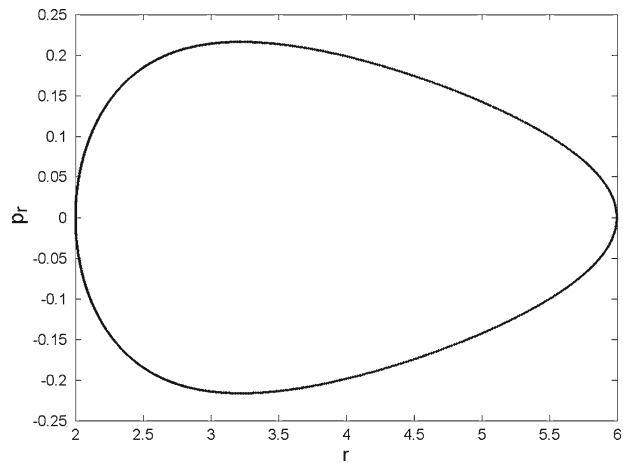

**Fig. 1** The Poincaré sections of the orbit A of a nonspinning ($S = 0$) particle around a maximal ($a = 1$) Kerr black hole, plotted in Boyer–Lindquist coordinates. The orbital parameters are $E = 0.8837$ and $J_z = 2.0667$, with $r_0 = 2.0$, $p^r = 0$, $\theta_0 = \pi/2$. The orbit is regular obviously

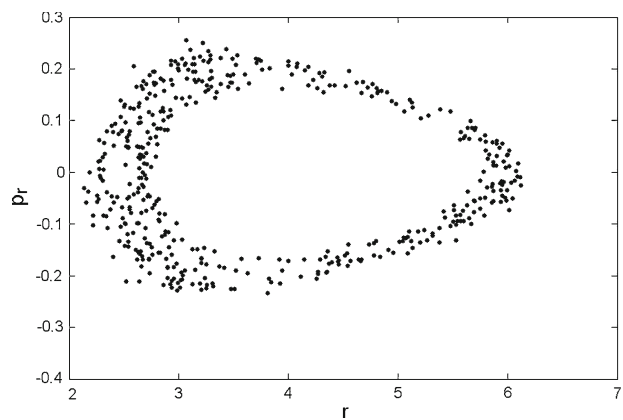

**Fig. 2** The Poincaré sections of the orbit B of a maximally spinning ($S = 1$) particle around a maximal ($a = 1$) Kerr black hole. The orbital parameters are $E = 0.8837$ and $J_z = 2.0667$, with $r_0 = 2.1$, $p^r = 0$, $\theta_0 = \pi/2$. The orbit B is likely chaotic

For the nonspinning case, because of existing four integral, so the system's freedom is four, and Poincaré sections are very suitable to detect chaos. We can find from the Fig. 1 that the orbit of a nonspinning particle (orbit A) is clearly integrability by the Poincaré sections.

For spinning particles, the degree of freedom increase to six. Poincaré sections are regular in higher dimensions space, but the projection to plane perhaps is stochastic for a nonchaotic orbit. So we think Poincaré section is not a good method to find

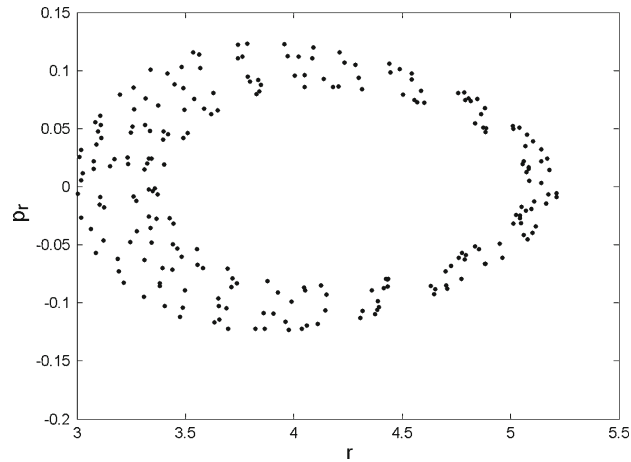

**Fig. 3** The Poincaré sections of the orbit C of a maximally spinning ($S = 1$) particle around a maximal ($a = 1$) Kerr black hole. The orbital parameters are $E = 0.8837$ and $J_z = 2.0667$, with $r_0 = 3.0$, $p^r = 0$, $\theta_0 = \pi/2$. The orbit C is a chaos mimic

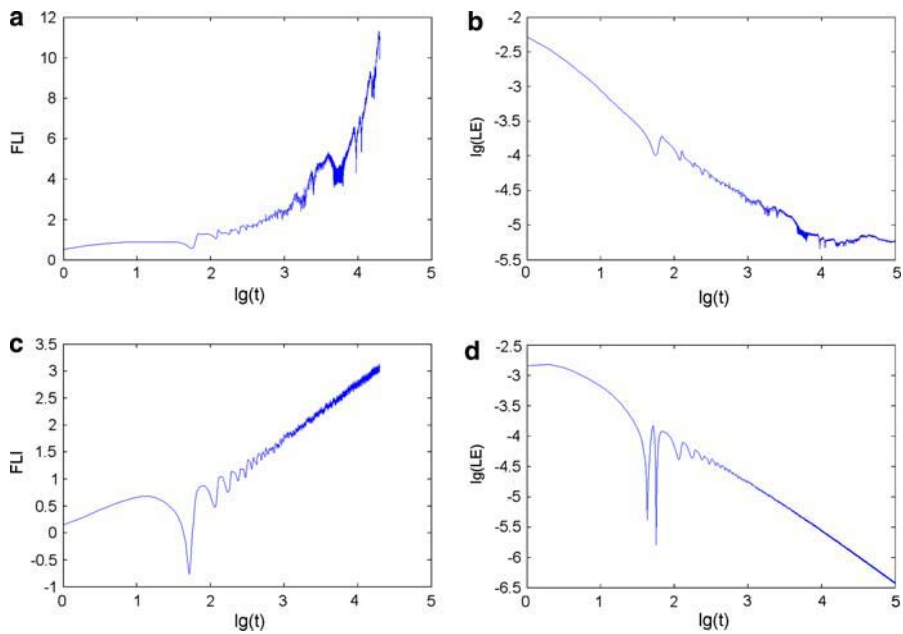

**Fig. 4 a** The FLI of orbit B. The chaos of the orbit B becomes explicit after a time span $10^3$. For this case, we just need to integrate to $2 \times 10^4$, and use only one renormalization. **b** The LE of orbit B. LE is convergent to $10^{-6}$ when time achieve to $10^5$. This show that the orbit B is chaotic. It has same conclusion with FLI. **c** The FLI of orbit C. The FLI increases in an algebraic law at time span $2 \times 10^4$, so this is a nonchaotic orbit. **d** The LE of orbit C. LE is down to zero when time achieve to $10^5$. This shows that the orbit C is nochaotic. It has same conclusion with FLI

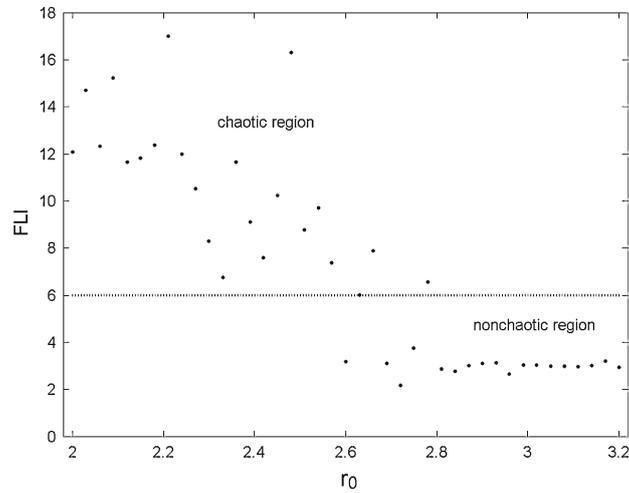

**Fig. 5** Scatter plot of initial radii $r_0$ versus FLI at $E = 0.8837$, $J_z = 2.0667$, $\theta_0 = \frac{\pi}{2}$, $S = 1.0$. While $r_0 \geq 2.8$, no chaos appear

chaos in our case. Increasing $S$ to 1, and fixing other parameters same with Fig. 1, we let $r_0 = 2.1$ (orbit B) in Fig. 2 and $r_0 = 3.0$ (orbit C) in Fig. 3, we find both them look likely irregular, but are both chaotic? By the help of FLI, we can find orbit B is chaotic, but orbit C is not, this is displayed in Fig. 4. In Fig. 4 we also calculate LE for comparing to FLI, and find they are consilient, but LE needs to cost more integration time.

## 6 Chaos in variable parameters

In this section, we investigate the relations between chaos and some vital parameters at certain energy and $z$ angular momentum, conclude: pericenter radii $r_p(p^r = 0$ section), polar angle $\theta_0$, Kerr rotation parameter $a$ and the spin magnitude $S$. When we discuss one of these parameters' influence to the dynamical system, must fix others. And in our all computing, we set coordinate time $t = 0$ at beginning. We typically compute to time $2 \times 10^4 M$, but if find chaos below the time $10^4$, we stop to calculate and save result.

### 6.1 Varying initial radii

In the Sect. 3, we point out that if the $r_0$ and $p_0^r$ are given, the apocenter and pericenter can be fixed, so in this section, we set all $p_0^r = 0$. From Eq. (14), we can deduce smaller $r_0(\approx r_p)$ means smaller $r_p$ and larger $r_a$, in the other word, larger eccentricity. The Fig. 5 exhibits the variety of FLIs with $r_0$ while other parameters invariable. The figure shows clearly that chaos disappear along as the increase of $r_0(\approx r_p)$, whereas exists two troubled dots, the $r_0 = 2.60$ and $r_0 = 2.78$. The FLI is primarily a function of initial radii $r_0$.

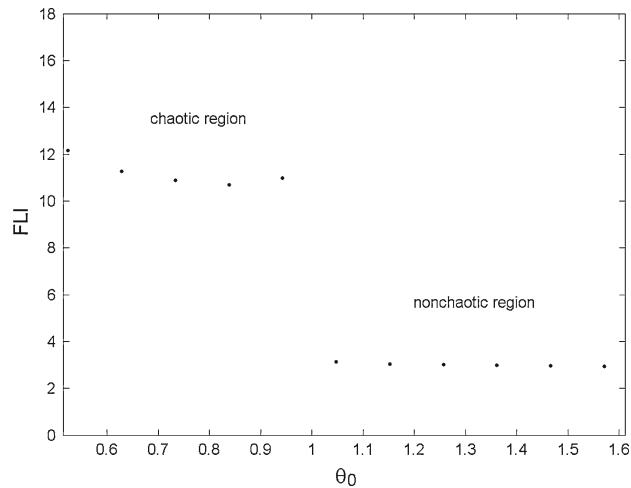

**Fig. 6** Scatter plot of initial polar angle $\theta_0$ versus FLI at $E = 0.8837$, $J_z = 2.0667$ and $S = 1.0$, $r_0 = 3.6$. Chaos appears at $\theta_0 = \pi/6$

### 6.2 Varying initial polar angles

The initial polar angle $\theta_0$ also has important influence on the character of the dynamical system. From Fig. 6, we can conclude that more small $\theta_0$ implies more easily for appearance of chaos. The other parameters are $E = 0.8837$, $J_z = 2.0667$, $S = 1.0$, $r_0 = 3.6$. At these conditions, the cutoff value of $\theta_0$ are $\pi/5$ (approximate), and below this critical value, we cannot find chaos.

### 6.3 Varying spin parameters

All we know, the spin magnitude $S$ is a most vital parameter to the occurrence of chaos, only while $S$ achieves a critical value, the system can become chaotic possibly. But, are all orbits chaotic while $S$ is bigger than the cutoff value if other parameters are invariable? Further more, is chaos stronger with $S$ increase while fixing other parameters? In the Fig. 7a, we do a rough scan which $S$ runs from 0.1 to 1.0 and $r_0$ is 2.0–3.0. This figure presents the chaotic orbits main locate large $S$. But more detailed calculations show that the bigger $S$ does not imply stronger chaos at all time, some times even cannot find chaos while other parameters are changeless. From Fig. 7b, we cannot find apparent rule between degree of chaos and $S$, but if $S$ is big enough, the system is chaotic always. The most strong chaos happens at $S = 0.95$, not 1.0. the other parameters in Fig. 7. are $E = 0.8837$, $J_z = 2.0667$, $a = 1.0$, $\theta_0 = \pi/2$.

If we consider all possible initial conditions, there is a critical $S$ for which chaos occurs. The nonchaotic orbits have a cutoff value of $S = 1$, i.e., they are not chaotic even in the extreme $S = 1$ limit. The cutoff value of $S$ was research by Hartl in 2003 [18], and did not find a physically realistic cutoff value. The detail was report in reference [18].

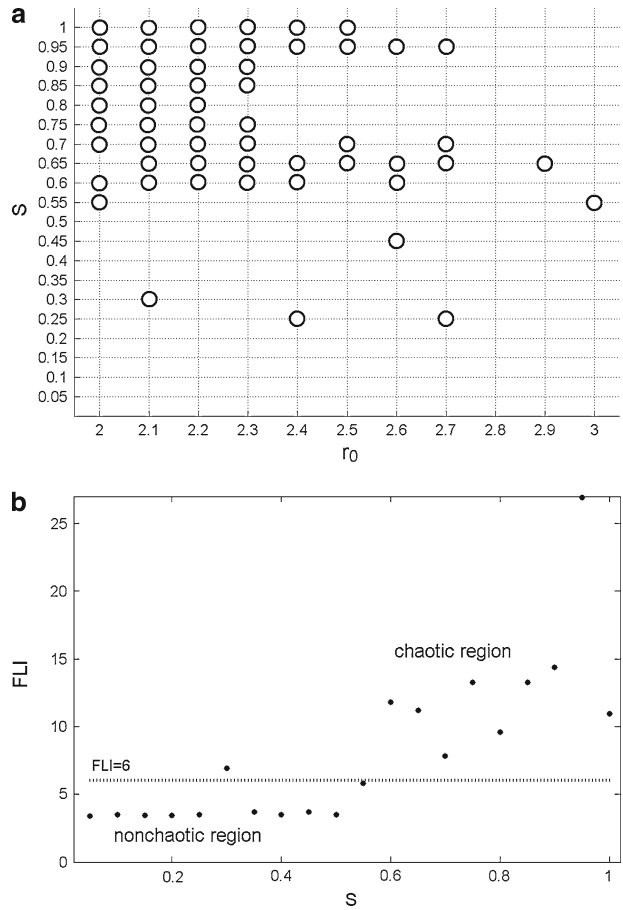

**Fig. 7** **a** A rough scan of spin parameter $S$ versus $r_0$ at $E = 0.8837$, $J_z = 2.0667$, $a = 1.0$, and $\theta_0 = \pi/2$. The symbol "O" presents the orbits are chaotic, else are nochaotic. **b** Scatter plot of spin parameter $S$ versus FLI at $E = 0.8837$, $J_z = 2.0667$, $r_0 = 2.0$, and $\theta_0 = \pi/2$. There is not apparent relation between chaos and $S$. The most strong chaos happens at $S = 0.95$

### 6.4 Varying Kerr parameter

The Kerr rotation parameter $a$ influences the Riemann–Christoffel curvature tensor directly, because of the coupling of the spin to the Riemann curvature, so the Kerr parameter $a$ should have effects on the spinning particles. We first do a rough scan which $a$ runs from 0.1 to 1.0 and $r_0$ from 3.0 to 5.0, and the conclusion is showed in Fig. 8a. When $a$ is small, the instable region is more big. It is obviously that chaos disappears at very big Kerr parameters. Also, we plot a part and detailed relation between the FLI and $a$ in Fig. 8b. We can find that the FLI is clear function about $a$. So the rotation of the center body perhaps can eliminate the effect of the spin of the particles, and counteract chaos. The other parameters in Fig. 8 are $E = 0.9328$, $J_z = 2.5667$, $S = 1$, $r_0 = 3.65$ and $\theta = \frac{\pi}{2}$.

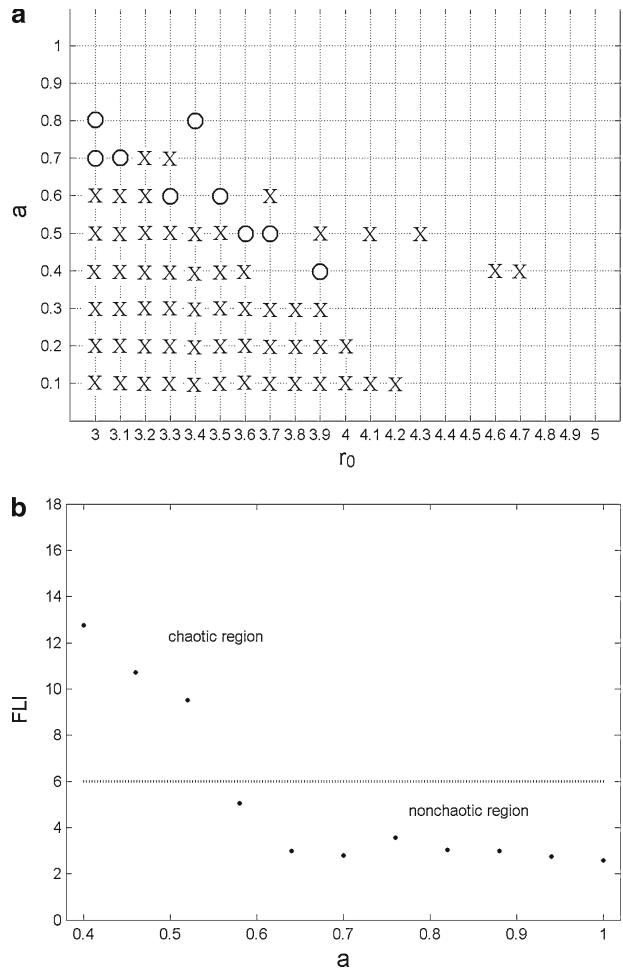

**Fig. 8** **a** A rough scan of Kerr parameter $a$ versus $r_0$ at $E = 0.9328$, $J_z = 2.5667$, $S = 1$, and the value of $\theta_0$ is same with Fig. 7. the Symbol "X" represents instable orbits. **b** Scatter plot of Kerr parameters $a$ versus FLI at $E = 0.9328$, $J_z = 2.5667$, $S = 1$, $r_0 = 3.65$, There is a clear relation between chaos and $a$. More small $a$, more easy to occur chaos

## 7 The exceptional dynamical character

There are some strange phase structure and asymmetrical about the equator plane for spinning particles with special initial conditions.

In the Fig. 9, the dynamical parameters are $E = 0.8837$, $J_z = 2.0667$, $r_0 = 3.5$, $\theta = \pi/2$, the Poincaré sections shows that it exists three intersectant orbits, so the system should locate at a equilibrium point, and it is nonchaotic. When $S$ decreases to 0.88, this phenomenon disappear. The Fig. 10 gives the section at $S = 0.89$, there are seem more intersectant orbits.

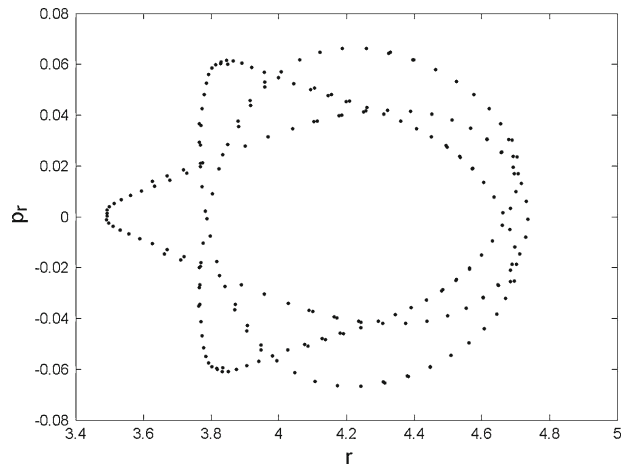

**Fig. 9** The Poincaré sections at parameters $E = 0.8837, J_z = 2.0667, r_0 = 3.5, \theta = \pi/2$, this clearly illustrate three intersectant orbits

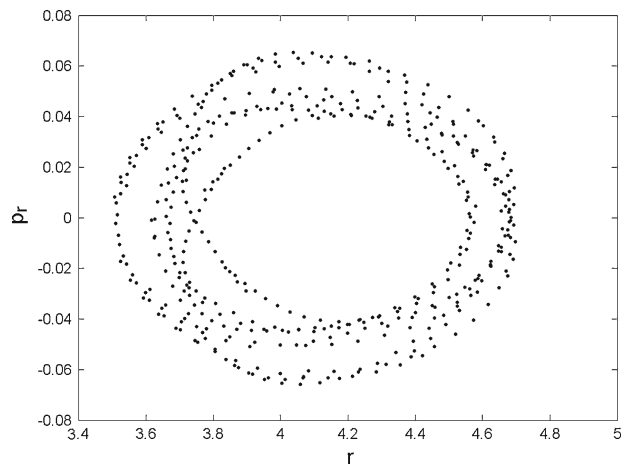

**Fig. 10** $S = 0.89$, others parameter are same with the Fig. 9. More orbits intersecting, and look likely stochastic but nonchaotic

Because of the symmetry of the Kerr metric, the spinless particles' orbits must be symmetry about the equator plane. In general, the spinning particles' trajectories also have symmetry about polar angle $\theta$ even when chaos happens, but for some special parameters, the orbits lose the symmetry, we give an example in Fig. 11. For studying this problem, we set $p^\theta = 0$ at fist, then deduce $p_0^r$. In Fig. 11, the parameters are $E = 0.9237, J_z = 2.8, S = 0.8, r_0 = 6.0, a = 1, p_0^\theta = 0$, and the orbit is nonchaotic. At the interval $0.64 \leq S \leq 0.89$, we can find the obvious asymmetry about equator plane.

Also, for spinless particles, if $p_0^\theta = 0, \theta_0 = \pi/2$, the movement should be astricted on equator plane. Obviously, if $S \neq 0$, it is impossible. But how the maximal inclination

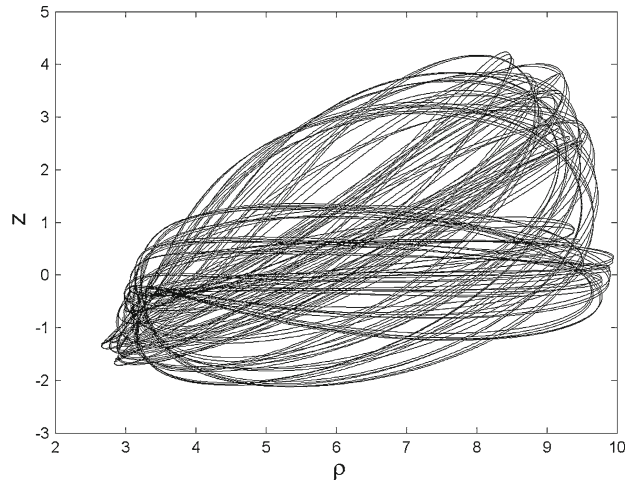

**Fig. 11** The trajectory projected to $z - \rho$ section, we find the loss of asymmetry about equator plane. The parameters are $E = 0.9237, J_z = 2.8, S = 0.8, r_0 = 6.0, a = 1, p_0^\theta = 0, \theta_0 = \pi/2$

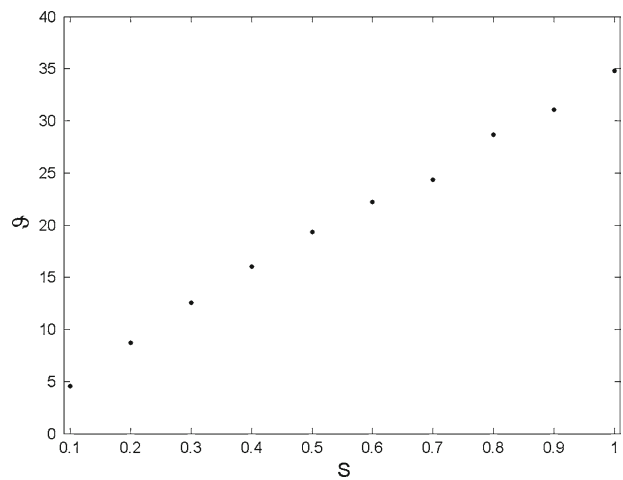

**Fig. 12** The function of the maximal inclination of orbit $\vartheta$ about $S$. The parameters are: $E = 0.9237, J_z = 2.8, r_0 = 6, a = 1, p_0^\theta = 0, \theta_0 = \pi/2$

of orbit ($\vartheta = \frac{\pi}{2} - \theta$) relates with spin parameter $S$? The Fig. 12 show us clearly that the $\vartheta_{\max}$ monotone increase with $S$. The other parameters are $E = 0.9237, J_z = 2.8, r_0 = 6, a = 1$. And all orbits are nonchaotic.

## 8 Conclusions

The Papapetrou equations, which model a spinning test particles moving in Kerr spacetime, exist many chaotic solutions for enough large $S$. And also, because of spinning, even if for a nonchaotic orbit, can exhibits some funny dynamical character depend on spin magnitude $S$. We detect chaos by means of FLI which was defined by Wu et al. in 2006, and emphasize that the Poincaré sections cannot be used to distinguish chaos. With the help of FLI, we give some significative relation between orbit chaos and dynamical parameters: pericenter, polar angles, spin and Kerr Parameters.

In our results, the rotation of central body can counteract the chaos effect aroused by spinning of particles, this is showed in Fig. 8. Though the orbit chaos need enough large spin magnitude, but the chaos is not a monotonic relation with $S$. The initial spin components $S^r$ and $S^\theta$ also have influence to chaos, this been study by Hartl in 2003 [18], and in this paper, we set the initial values $S^r = S^\theta = 0.1S$. Furthermore, for nonchaotic orbits, the maximal inclinations are monotone increasing with spin parameter $S$. Some surprise dynamical structures appear at given parameters. For example, while $E = 0.8837, J_z = 2.0667, r_0 = 3.5, \theta = \pi/2$, Poincaré sections show that there are three orbits superpose together (Fig. 9). Sometimes, because of spin, the orbits are asymmetric about the equator plane.

Notwithstanding Hartl [17,18] pointed out the orbits are nonchaotic for physically realistic values of $S$ (satisfying $S \ll 1$), but the model is still interesting. And in this paper, we do not consider the gravitational wave radicalization form the model, and would consider it in future.

**Acknowledgments** The author thanks to Professor Xinhao Liao for his useful discussions. And, Professor Xin Wu also gave some good advice for this paper.